\documentclass[DIV=calc, paper=a4, fontsize=11pt, twocolumn]{scrartcl}	 
\usepackage{ifpdf}
\ifpdf
\pdfoutput=1
\fi

\usepackage[english]{babel} 
\usepackage{amsmath,amsfonts,amsthm} 
\usepackage[svgnames]{xcolor} 
\usepackage[hang, small,labelfont=bf,up,textfont=it,up]{caption} 
\usepackage{booktabs} 
\usepackage{fix-cm}	 
\usepackage{overcite}
\usepackage{graphicx}
\usepackage{float}
\usepackage{eurosym}
\ifpdf
\usepackage[final,tracking=true,kerning=true,spacing=true,factor=1100,stretch=10,shrink=10,protrusion=true,expansion=true]{microtype}

\usepackage[colorlinks,bookmarks=true]{hyperref}

\fi

\usepackage{sectsty} 
\allsectionsfont{\usefont{OT1}{phv}{b}{n}} 
\usepackage{fancyhdr} 
\usepackage{lastpage} 

\newcommand*{\skippingparagraph}{\par\vspace{\baselineskip}}
\def\CC{{C\nolinebreak[4]\hspace{-.05em}\raisebox{.4ex}{\tiny\textbf{++}}\hspace{.15em}}}

\usepackage{lettrine} 
\newcommand{\initial}[1]{ 
\lettrine[lines=3,lhang=0.3,nindent=0em]{
\color{DarkGoldenrod}
{\textsf{#1}}}{}}

\usepackage{titling} 
\newcommand{\HorRule}{\color{DarkGoldenrod} \rule{\linewidth}{1pt}} 
\pretitle{\vspace{-30pt} \begin{flushleft} \HorRule \fontsize{50}{50} \usefont{OT1}{phv}{b}{n} \color{DarkRed} \selectfont} 
\title{Fast Compression Method for Medical Images on the Web} 
\posttitle{\par\end{flushleft}\vskip 0.5em} 
\preauthor{\begin{flushleft}\large \lineskip 0.5em \usefont{OT1}{phv}{b}{sl} \color{DarkRed}} 
\author{Dr. Bas Hulsken, } 
\postauthor{\footnotesize \usefont{OT1}{phv}{m}{sl} \color{Black} 
	CTO Philips Digital \& Computational Pathology
\par\end{flushleft}\HorRule} 

\date{May 2020}

\begin{document}

\maketitle 

\thispagestyle{fancy} 


\initial{T}\textbf{he need for fast diagnostic image viewing in zero footprint web applications and the ever increasing image sizes for new modalities such as digital pathology have painfully brought to light that the currently available image compression methods fall short. JPEG2000 delivers the image quality required for medical grade viewing, but is supported on fewer than 10\% of desktop web browsers installed today (\ifpdf \href{https://caniuse.com/\#feat=jpeg2000}{caniuse.com}\else caniuse.com \fi) and even then it does not support the high bit depth images required by medical applications. JPEG2000's high computational complexity and inability to do fast compression and viewing of images undoubtedly contributed to its lack of adoption. The venerable JPEG standard is supported in all installed web browsers today, and allows for fast viewing and compression, but it cannot provide medical grade image quality, lossless compression, or high bit depths. To remedy the situation medical image web applications need to take full control of the image path by implementing the image decompression in the application itself, instead of relying on web browser implementations. This will give users and manufacturers the confidence that medical images will be displayed in their full intended fidelity! This paper introduces a simple, fast, yet efficient image compression method that can be implemented in zero footprint web applications to provide fast medical grade image viewing on today's web browser installed base, on the premise and in the cloud. Due to its efficiency it is very suitable for fast and affordable compression and viewing of very large images and is used as such by Philips in the iSyntax file format in use in its digital pathology products today.}

\ifpdf
\hypersetup{pdfkeywords={Digital Pathology, Whole Slide Images, Image Compression },urlcolor={blue},citecolor={magenta} }
\fi


\section{Introduction}

While other industries have led the way in moving from traditional locally installed software to web and cloud applications, healthcare is not far behind. Today many medical software applications are being replaced by web and mobile applications that bring benefits such as reduced cost, increased reliability, ubiquitous access and shorter upgrade cycles. 

An important requirement for many medical software applications, and in particular for diagnostic imaging applications, is the ability to display and interact with high-fidelity images. And there lies a problem: as can be clearly deduced from the poor support for JPEG2000 -- less than 10\% of today's installed desktop web browsers support it, 20 years after the introduction of the standard\cite{caniuse:jpeg2000} -- as well as from the complete absence of support for high bit depth images, web browsers have not catered to the needs of medical imaging applications.

Even in the unlikely event that web browser manufactures will enable a standardized high fidelity image path suitable for medical diagnostic imaging on short notice, it will take many years before these features are readily available in the installed base of web browsers. Implementing image decompression and display in the web application itself is therefore the only solution for applications that need reliable, high fidelity image support in web applications today. Various open-source libraries are available to do just that (\ifpdf \href{https://www.npmjs.com/package/image-js}{image-js} \else image-js \fi supports many image formats such as 16-bit PNG, TIFF, \ifpdf \href{https://github.com/kripken/j2k.js}{j2k.js} \else j2k.js \fi is a JavaScript JPEG2000 decoder). The problem here is that there is no image compression method that is powerful, yet simple enough to yield the image quality and versatility required by medical imaging applications at acceptable speeds when implemented in web applications with JavaScript.

The compression method described in this paper is the result of the work done at Philips Digital and Computational Pathology to solve this challenge for the giga-pixel sized whole slide images that need to be compressed at rates of hundreds of mega-pixels per second into a format that allows fast panning and zooming from zero footprint web applications.

\section{Required Functionality}

The three key requirements for a practical image compression method for zero footprint web applications for medical imaging are: \textbf{full control} over displayed pixels, \textbf{unrestricted image quality} by not imposing quality constraints in the image compression method itself, and \textbf{fast image delivery} through very fast decompression at good compression ratios.

\textbf{Full control} over the display pixels can practically only be attained by implementing the entire image rendering path, including decompression and post-processing, in web app itself. The alternative of using the widely supported lossless PNG or BMP image formats would restrict image quality (8-bit  gray scale or 24-bit true color) and prevent fast image delivery (poor or no compression respectively). When implementing image decompression and post processing in the web app, computationally expensive operations can be performed in WebGL, which is supported on more than 99\% of the desktop installed base, and more than 97\% of mobile installed base\cite{caniuse:webgl} today.

\textbf{Unrestricted image quality} requires that the compression method does not put additional limitations on the image quality, such as a limitation to 8-bits per gray scale or color channel, a forced resolution reduction for color information or a limitation on supported color spaces and the number of color channels. The widely supported JPEG standard has all these limitations which means that relying on it for web based medical applications would result in a reduction of image quality with respect to existing workstation applications.

\textbf{Fast image delivery} requires a pragmatic trade-off between computational complexity of the decompression algorithm (to allow for fast rendering speeds even in JavaScript) and compression ratio (to allow for fast downloads). The JPEG2000 standard delivers on unrestricted image quality, yet it is very poorly supported with less than 10\% of the current desktop browser install base\cite{caniuse:jpeg2000}. This lack of adoption, 20 years after introduction, is for a considerable part related to JPEG2000's high computational complexity. That aspect makes an implementation of a JPEG2000 decompressor in the web app itself unfeasible, in particular because the slowest part of JPEG2000 compression is the embedded bit plane and entropy coder (EBCOT) which cannot be accelerated by using WebGL; up to 82\% of execution time for CPU implementations is spent on EBCOT, and up to 93\% for GPU implementations\cite{Ciznicki:Benchmarking,Moccagatta:Computational}.

\skippingparagraph
We can conclude that there is no image compression method available today that fully delivers on the needs of zero footprint web applications for medical imaging. JPEG cannot deliver an image quality that is on par with today's medical imaging software running on workstations. PNG and BMP impact fast image delivery due to poor compression ratios, and JPEG2000 has a too small an installed base; a problem which cannot be addressed by directly implementing JPEG2000 decompression in the web app itself because due to its computational complexity it would not allow for an acceptable image delivery speed.

\section{Proposed Compression Method}

The compression method proposed in this paper delivers on all requirements for a medical zero footprint app and is a pragmatic combination and enhancement of well-known methods from the field of image compression. Consistently, when design decisions required trade-offs to be made, decisions favored low computational complexity, ease of implementation and reuse of existing algorithms over compression ratio. No compromises were made however with respect to features required by medical imaging: the compression method proposed here allows for arbitrarily high bit-depths, unlimited number of channels, lossless and lossy compression and progressive decompression in terms of resolution as well as quality.

\begin{figure}[t]
  \noindent\centerline{\includegraphics[width=0.5\textwidth]{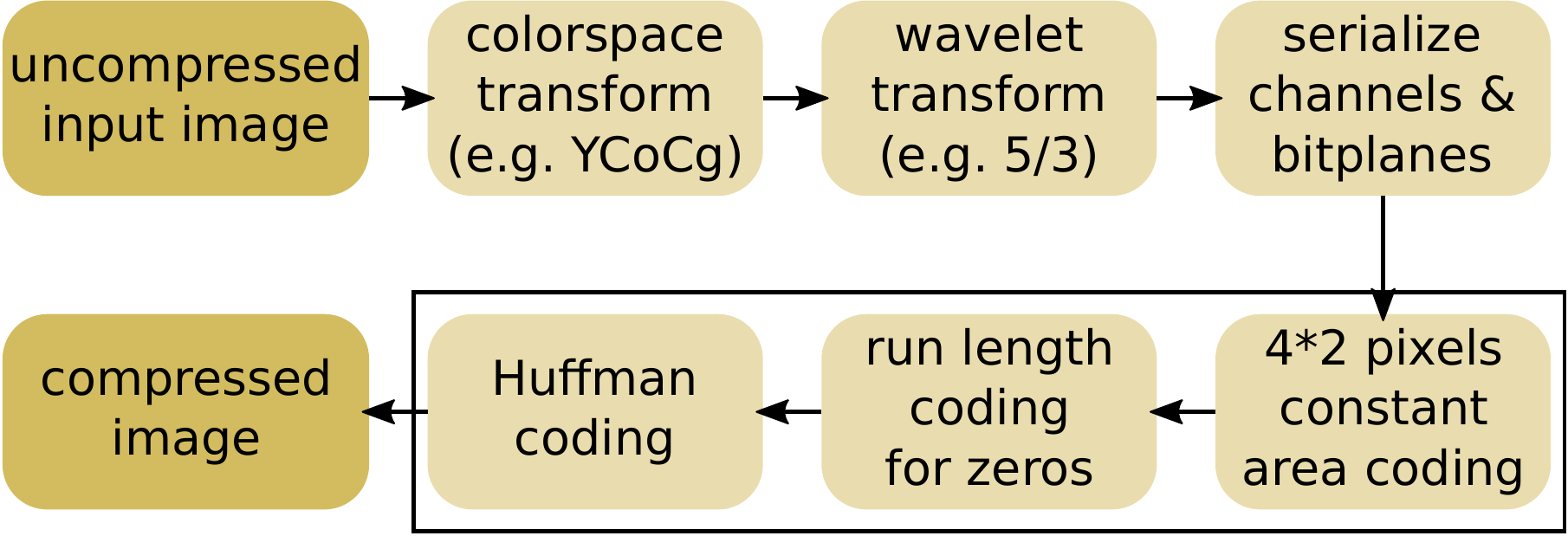}}
	\caption{Schematic representation of the image compression principle proposed in this paper. The functional blocks enclosed in the black rectangle are well-known algorithms that are modified or enhanced as part of this work. The other functional blocks are well-known algorithms for which of the shelve implementations were used for the results published in this paper.}
  \label{fig:compression_principle}
\end{figure}

A functional decomposition of the image compression pipeline is shown in Figure~\ref{fig:compression_principle}. The concept can be easiest understood as a generalization of constant area bitplane coding (CAC)\cite{Marques:Practical} combined with a run length and Huffman entropy coder applied to wavelet transformed bit-planar data. The first 3 blocks in the compression pipeline are standard off-the-shelf transformations: an (optional) colorspace transformation, a recursive wavelet transformation and a serialization into consecutive bitplanes. A wide variety of implementations are readily available in software libraries and image processing toolkits such as Matlab\cite{Mathworks:Matlab}, which has been used to create the reference implementation\cite{reference:iSyntax} of the compression method that is provided as accompanying information with this paper. The novelty in the compression method proposed here lies in the particular combination and modification of the subsequent 3 functional blocks.

\begin{figure}[t]
  \noindent\centerline{\includegraphics[width=0.5\textwidth]{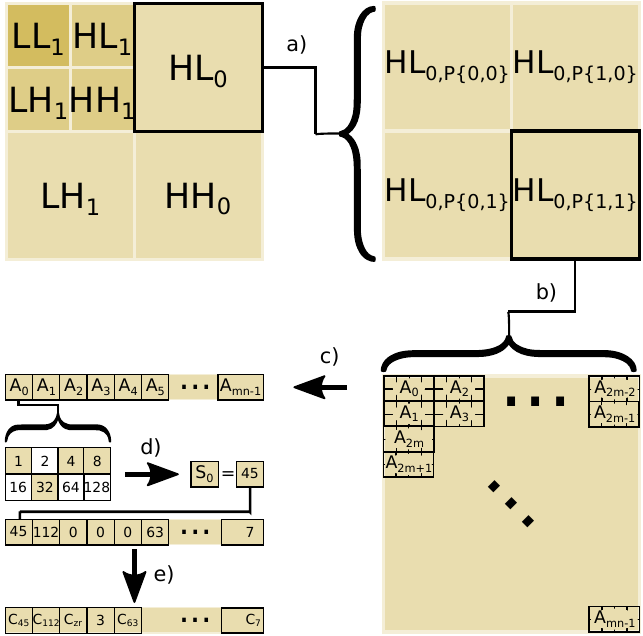}}
	\caption{Schematic representation of the serialization, constant area coding and Huffman + run length coding functional blocks of the compression pipeline shown in Figure~\ref{fig:compression_principle}. After the (recursive) wavelet transform, the wavelet coefficients will be transformed and partitioned into a bit-planar format (\textbf{a}). Each of the partitions will be subdivided into $4*2$ areas $A_i$ (\textbf{b}). These areas are then serialized in snake order (\textbf{c}) and assigned a unique symbol $S_i$ in the range $[0,255]$ by mapping each pixel in the $4*2$ area to a bit of the symbol in a clockwise fashion (\textbf{d}). Finally, the symbols are Huffman coded, zero runs are coded with special codeword $C_{zr}$ followed by a run length counter (\textbf{e}). $C_{zr}$ can double as code for a normal symbol (counter is 0) to ensure fast 8-bit Huffman codes.}
  \label{fig:compression_algorithm}
\end{figure}

In standard constant area bitplane coding a binary image or an individual bitplane of a non-binary image, is divided into multiple $m*n$ pixel areas that are subsequently assigned one of 3 possible fixed-length codewords, coding an all-black, all-white, or a mixed region. In the case of a mixed region, after the codeword, the individual pixel values are stored. The core compression principle used here is the extension of this standard constant area bitplane coding from only 3 fixed-length codewords, to codewords for each combination of pixels in the $m*n$ area occurring in the image. Using a $4*2$ area ensures a fast algorithm as that would allow an area (having 256 possible different configurations) to be represented by a symbol of a single byte. This mapping of an area $A_0$ to its corresponding unique symbol $S_0$ is show in step \textbf{d)} in Figure~\ref{fig:compression_algorithm}. After mapping the serialized areas to their corresponding symbols, a Huffman code\cite{Huffman:Method} is used to compress the stream of symbols. Due to the  wavelet transform, in particular bitplanes of the high-pass wavelet coefficients will contain significant all zero areas, which means that an improved compression ratio can be attained by combining the Huffman coding with run length coding for the zero symbols.

The choice of a recursive wavelet transformation combined with a bit plane sequential compression allows for the progressive decompression of very large images at different scales and quality levels, similar to the functionality JPEG2000 provides. This is beneficial for fast zooming and panning through very large images such as digital pathology whole slide images. Providing a similar functionality with JPEG compressed images requires a redundant pyramid of lower resolution versions of the same image, leading to a 30\% increase in file size.

The choice for a simple bitplane sequential compression method enables an arbitrary number of channels and an arbitrary bitdepth for image quality without constraints; again, similar to what JPEG2000 provides. To create lossy images, the same image quantization methods as in JPEG2000 can be used.

Finally: the image compression method proposed here is a simple easy to understand lossless, or constant quality lossy compressor. The simplicity and explainability is a benefit in itself in the field of medical image compression, in particular when utilizing artificial intelligence for image interpretation, which can be sensitive to subtle distortions and image compression artifacts.

\section{Implementation}

\begin{figure}[t]
  \noindent\centerline{\includegraphics[width=0.5\textwidth]{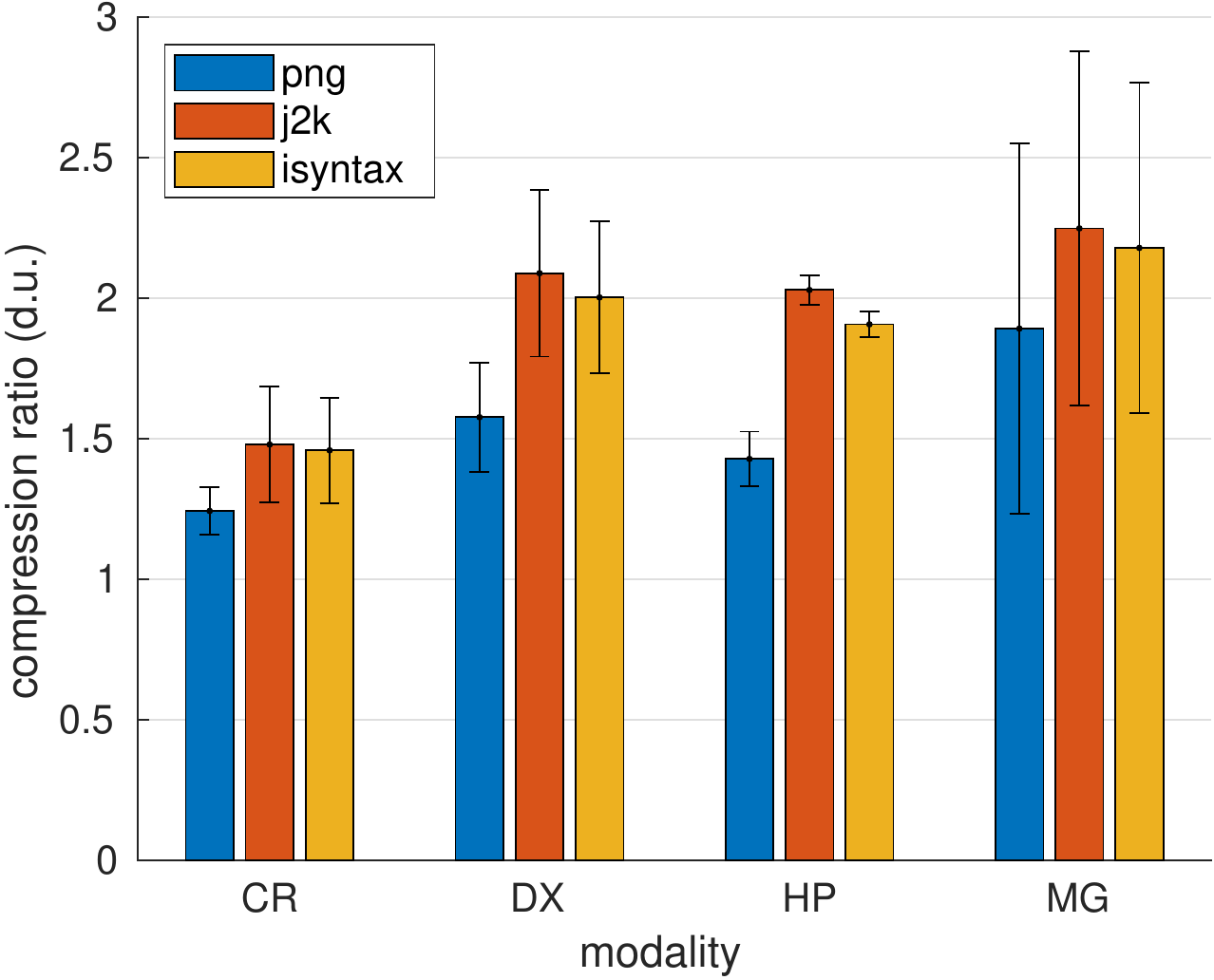}}
	\caption{Lossless compression ratio of iSyntax (yellow) versus JPEG2000 (orange) and PNG (blue) for computed radiography\cite{CPTAC:PDA} (CR), digital x-ray\cite{CPTAC:PDA,CPTAC:LUAD,CPTAC:LSCC,TCGA:BLCA,TCGA:GBM} (DX), histopathology visible light (HP) and mammography\cite{TCIA:VICTRE,TCGA:BRCA,TCIA:CBISDDSM} (MG) images. On average the iSyntax compression proposed in this paper (using the MATLAB reference implementation included in the accompanying information) results in 3.4\% larger files than with JPEG2000 compression (using Jasper), and 17.7\% smaller files than with PNG (using MATLAB imwrite) compression.}
  \label{fig:compressionratios}
\end{figure}

The compression ratio benchmark in Figure~\ref{fig:compressionratios} was generated using a reference implementation of the compression method presented in this paper written in MATLAB. Said reference code is provided as accompanying information with this publication.

The color space transformation, wavelet transformation and bitplane serialization routines that are part of the compression pipeline are standard, and are not discussed here any further than to mention that MATLAB's 5/3 Le Gall integer lifting wavelet transformation was used, and that wavelet data was partitioned in $128*128$ blocks before compression. The color space transform, the wavelet transformation and the bitplane serialization all allow for very fast implementations using SSE, GPUs, ASICs, or FPGAs. The implementation details of such fast implementations are beyond the scope of this paper, yet the benchmarks shown in Figures~\ref{fig:compressionspeedssmall} and~\ref{fig:compressionspeedslarge} are attained with highly optimized parallel \CC code and employ SSE2 instructions available on any modern x86 CPU.

The serialization of a single channel of wavelet coefficients (single color and coefficient type LL, HL, LH or HH) into $4*2$ binary areas $A_k$ is performed using a snake ordering and a signed magnitude representation (SMR) of the coefficients, where the first serialized bitplane is that of the sign bit, followed by the magnitude bits, most significant bit down to the least significant bit.

The linear ordering of the $4x2$ areas in a coefficient block of width $w$, height $h$ and bit-depth $d$ is determined by a mapping function that maps position $k$ in the serialized stream of binary areas $A_k$ to a coordinate $(i,j,b)$ giving the top left corner coordinate of that area and the bitplane it represents. This (non-linear) transformation $T: \mathbb{N}_0 \rightarrow \mathbb{N}_0^3$ is given by:
\begin{equation}
	T(k)=
	\begin{pmatrix} i \\ j \\ b \end{pmatrix} =
	\begin{pmatrix}
		k \% 2 +2 \lfloor (k \%  (wh/8))/2w \rfloor \\
		4( \lfloor k/2 \rfloor \%(w/4)) \\
		\lfloor 8k/wh \rfloor + d - 1 
	\end{pmatrix}
	\label{eqn:areacoord}
\end{equation}
with $\%$ the modulo operator. Step \textbf{c)} in Figure~\ref{fig:compression_algorithm} illustrates this mapping. The transformation function of a binary area $A_k$ into an 8-bit symbol $S_k$, with $B_{(m,n,b)}$ the binary value of bit $b$ of the coefficient at position $(m,n)$ in the coefficient plane, is as follows: 
\begin{equation}
	S_k = \sum_{m=i}^{i+1}\sum_{n=j}^{j+3} B_{(m,n,b)} 2^{n-j+4(m-i)}
	\label{eqn:serialize}
\end{equation}
with $(i,j,b)$ the coordinate of the top left corner of binary area $A_k$ as given by Equation~\ref{eqn:areacoord}. This transformation is illustrated by step \textbf{d)} in Figure~\ref{fig:compression_algorithm}, and is the first step of the enhanced constant area coding (CAC) step. The serialization of the coefficient data described by these two equations can be implemented easily using vectorization techniques, allowing for fast implementation on any CPU with a Single Instruction Multiple Data (SIMD) instruction set, or on a GPU or FPGA.

The second step of the constant area coding comprises assigning a codeword to each of the symbols $S_k$ using Huffman coding. But before the Huffman coding two additional steps are performed to improve the compression ratio. Firstly, for the bitplanes in the coefficient block which are all zero, the entire consecutive block of symbols $S_k$ (all having a value of zero) for that bitplane are removed from the stream. A simple bitmask prepending the compressed block is used to indicate bitplanes that were removed in this manner. Secondly, zero-runs, defined as two or more consecutive zeros in the symbol stream, are identified and replaced by a zero-run symbol followed by a value describing the length of the zero-run. The least frequently occurring symbol in the set of $S_k$ symbols representing the coefficient block is chosen as the zero-run symbol. However, if that symbol also occurs in the coefficient block, then the original symbol is stored as a zero-run of length zero. It would not be efficient to store the zero-run length with a bit depth large enough to store the length of the longest zero-run in the stream; rather a smaller number of bits is chosen, and longer runs can be stored by a repeated sequence of zero-run symbols and counters. To determine the optimal bit depth of the zero-run counter one needs to solve the minimization problem in equation~\ref{eqn:counterbits} where $w$ is the optimal counter bit depth for a block of data of $2^b$ bytes in length, where $F_i$ is the frequency of occurrence of zero runs with a length that require $i$ bits to store, and $l$ is the length in bits of the zero-run symbol's corresponding Huffman codeword.
\begin{equation}
	\min_{w \in [2,b]} \sum_{i=1}^{bits}{\lceil F_i/w \rceil \cdot (w+l)}
	\label{eqn:counterbits}
\end{equation}
Now that the empty bitplanes have been removed and zero-runs have been replaced by zero-run symbols and counters, the actual Huffman coding takes place. The Huffman code is created as normal, where the frequency of the symbols is determined taking into account the number of zero-runs. After the determination of the Huffman code, the symbol stream is Huffman coded by replacing each symbol by its corresponding codeword; the zero-run counters are not Huffman coded but stored with big endian bit order. The Huffman and zero-run coding are illustrated as step \textbf{e)} in Figure~\ref{fig:compression_algorithm}.
\begin{figure}[t]
  \noindent\centerline{\includegraphics[width=0.5\textwidth]{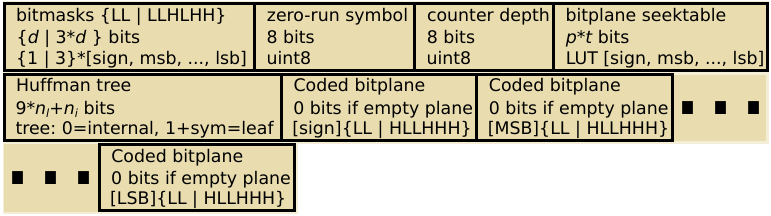}}
	\caption{Fields in the compressed block header; the order is left to right, top to bottom. The fields are not byte aligned. A block can contain either LL coefficients, or multiplexed LH, HL and HH coefficients of bit depth $d$. The seektable consists of pointers into the bitstream, of depth $t=\lfloor log_2(p*w*h/8) \rfloor + 5$ bits. A single $d$ bits bitmask per coefficient type is stored. The Huffman codes are stored as a recursive tree with a 0 representing an internal node, and a 1 followed by the 8-bit symbol value a leaf node.}
  \label{fig:blockheader}
\end{figure}
The compression of the coefficient block is now complete, the composition of the compressed block is shown in Figure~\ref{fig:blockheader}.

Decompression is simply traversing the pipeline show in Figure~\ref{fig:compression_principle} in reverse. Because the compression coefficients are stored in separate blocks for each scale, it is possible to reconstruct the image at any resolution level without expensive down-scaling by simply discarding the high-pass coefficients representing the higher resolutions. Because each compressed block contains a seek table pointing to individual bitplanes, which are ordered in decreasing significance in the block, it is possible to reconstruct an image by discarding an arbitrary number of the least significant bits for the block, allowing for fast reconstruction and transmission of the image data at a lower quality without expensive decompression and recompression.

\section{Benchmarking}

The compression performance of the method introduced in this paper is compared to both JPEG2000 and PNG in Figure~\ref{fig:compressionratios}. We compare only the lossless performance here, since it is the most straightforward comparison since the image quality is identical for each of the methods. We have chosen to compare against JPEG2000 and PNG since these are the only readily available lossless formats. A representative set of radiology images (computed radiology, digital x-ray, and mammograms, abbreviated as CR, DX, MG respectively) was taken from The Cancer Imaging Archive\cite{Clark:TCIA}. The pathology images (histopathology abbreviated as HP) were uncompressed scans of a set of histopathology slides using the Philips scanner at 9 bits per color channel.

\begin{figure}[t]
  \noindent\centerline{\includegraphics[width=0.5\textwidth]{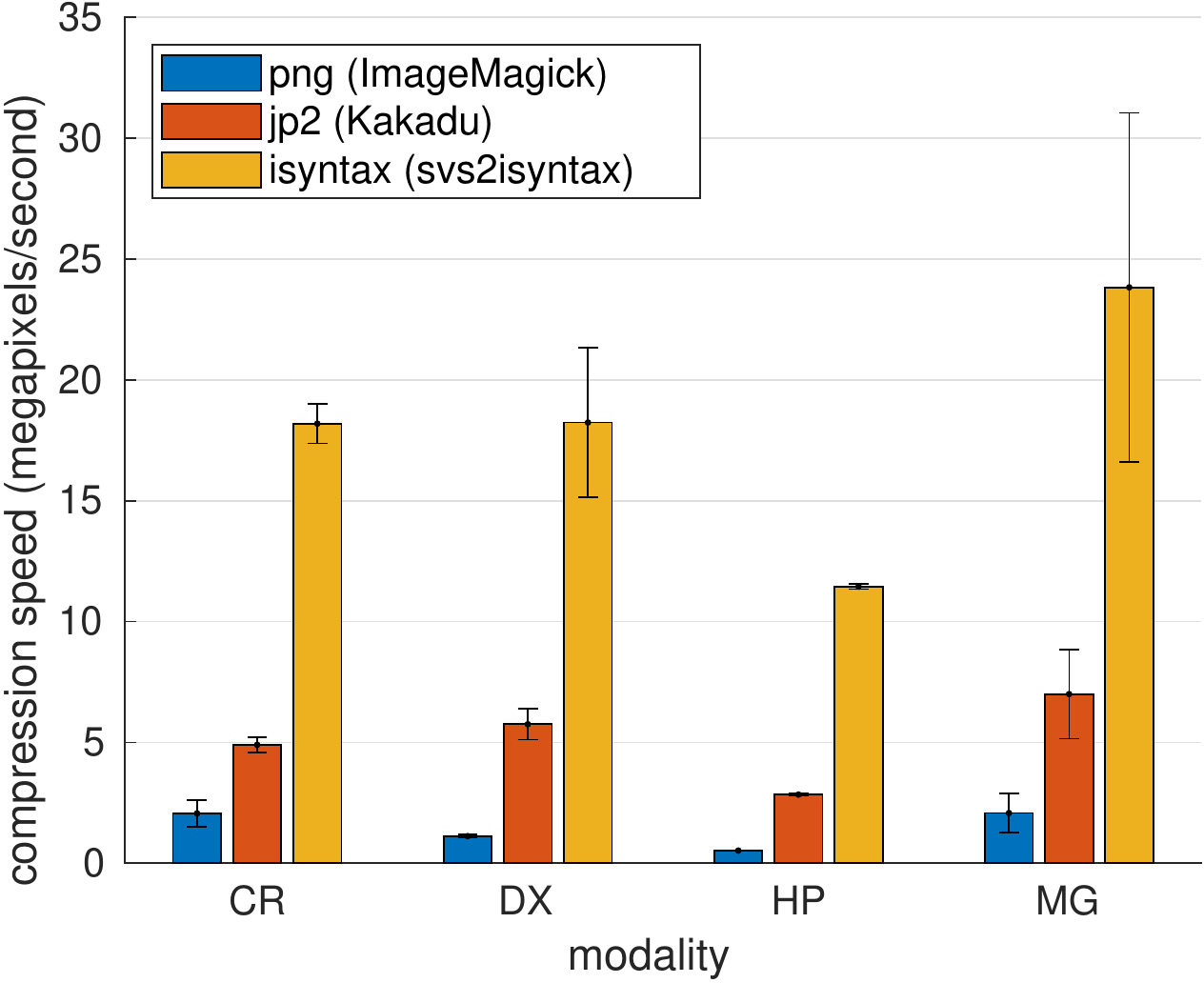}}
	\caption{Lossless compression performance of iSyntax (purple) versus JPEG2000 (orange) and PNG (blue) for the same images in Figure~\ref{fig:compressionratios}. Kakadu's kdu\_compress v7.2.10 was used for the JPEG2000 compression, ImageMagick 6.9.10 with libpng 1.6.37 was used for the PNG compression. In terms of speed, the iSyntax implementation is 250\% faster than Kakadu's JPEG2000 and 1140\% faster than ImageMagick's PNG compression. For a fair comparison of the computational complexity of the compression method the benchmarks were run in single threaded mode.}
  \label{fig:compressionspeedssmall}
\end{figure}

\begin{figure}[t]
  \noindent\centerline{\includegraphics[width=0.5\textwidth]{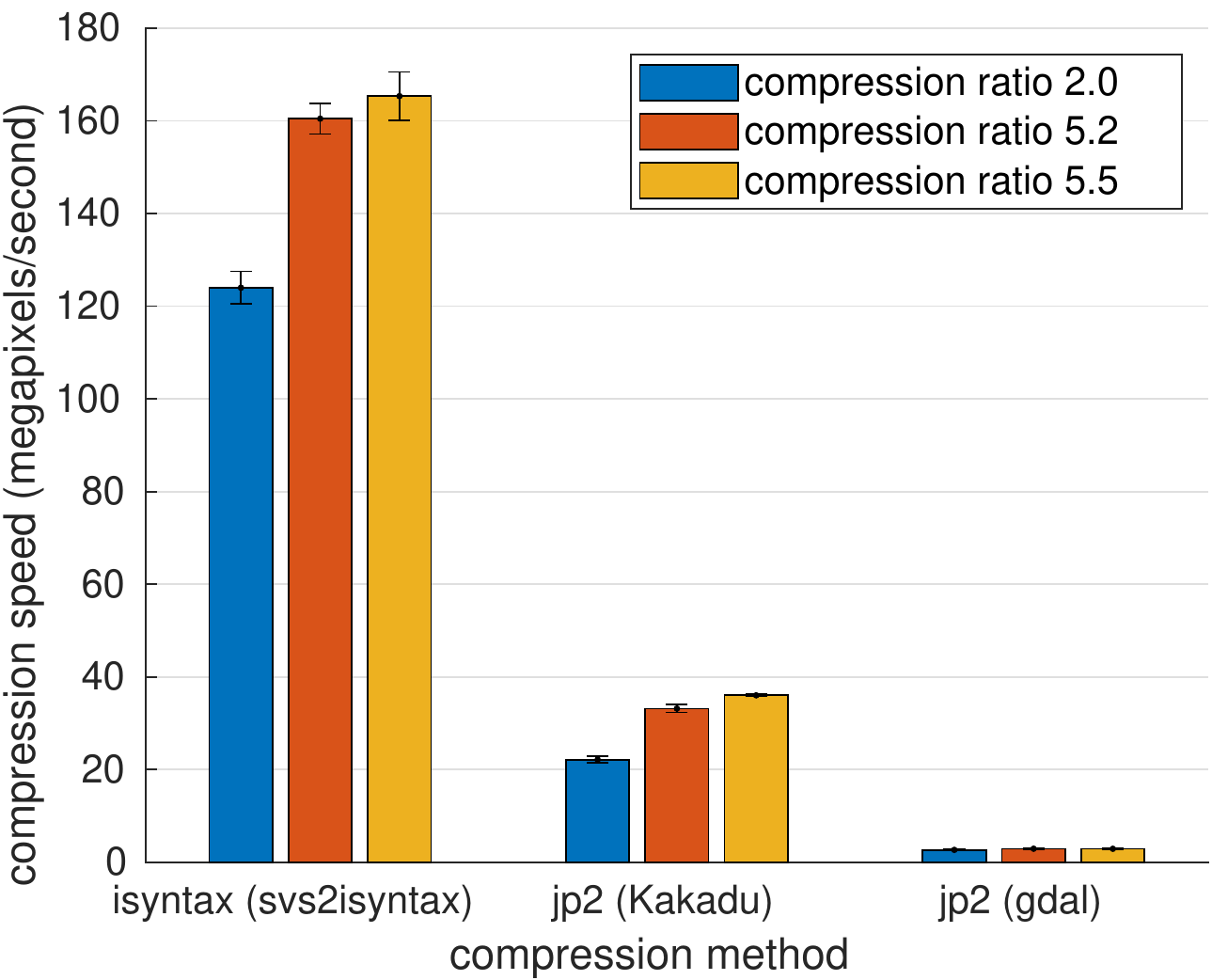}}
	\caption{Lossless compression performance of iSyntax versus JPEG2000 with Kakadu's kdu\_compress and openjpeg 2.3 via gdal\_translate. To estimate a realistic real-word compression performance, this time the compression was run with multi-threading enabled. For lossless compression (blue) iSyntax is 460\% faster than Kakadu's JPEG2000 implementation, and 4596\% faster than openjpeg 2.3 used via gdal\_translate of the Geospatial Data Abstraction Library (GDAL) 2.3.2, which is disadvantaged by its inability to use multiple threads, yet is the only free compression tool available for creating very large JPEG2000 images. Both iSyntax and JPEG2000 compression are faster when enabling a slight lossy compression of a ratio of around 5.}
  \label{fig:compressionspeedslarge}
\end{figure}

To compare the computational complexity of the compression method for compression, the speed in megapixels per second was measured for the same images and compression methods with optimized compiled code run on a normal desktop PC (AMD Ryzen 5 3600 6-Core Processor, sub \EUR{200} CPU) running Fedora Linux version 31. The results are shown in Figure~\ref{fig:compressionspeedssmall}. The low computational complexity of iSyntax pays off in significantly higher speeds, yet with small images the speed is limited by the time it takes to open and close the image and start the executable. For larger images these overheads are insignificant, and the raw speed of the compression method becomes apparent. In Figure~\ref{fig:compressionspeedslarge} the compression speeds for 100-megapixel pathology whole slide images (27-bit RGB) are shown.

\begin{figure}[t]
  \noindent\centerline{\includegraphics[width=0.5\textwidth]{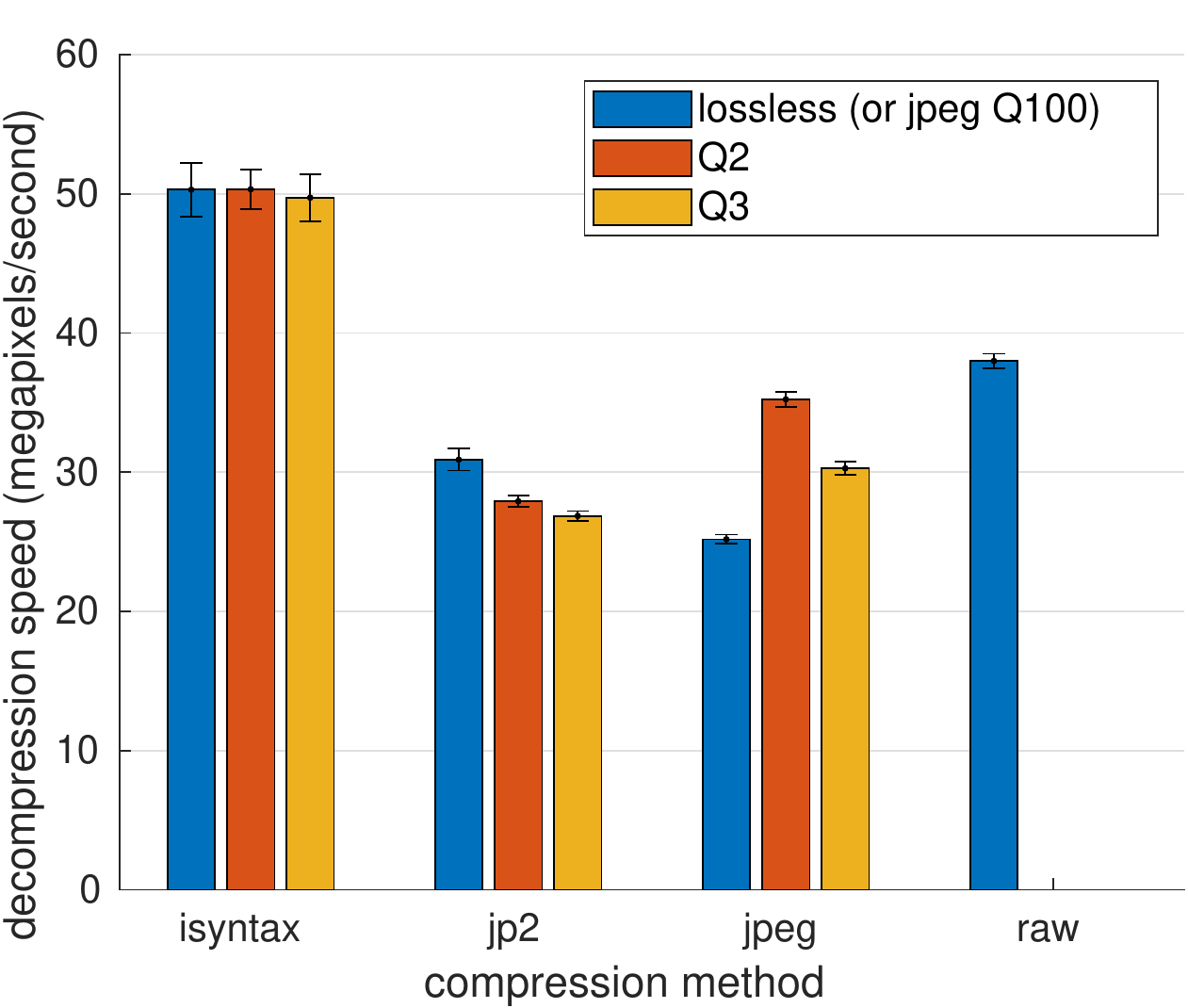}}
	\caption{Decompression performance of iSyntax at several quality levels versus JPEG2000 (openjpeg 2.3) JPEG (openslide 3.4.1 with libjpeg-turbo 2.0.2 and libtiff 4.0.10) uncompressed RAW (openslide 3.4.1 with libtiff 4.0.10). Also, for decompression the low complexity of iSyntax pays off, yet with smaller differences than the compression.}
  \label{fig:tileextractspeed}
\end{figure}

\section{Conclusions}

The compression method introduced in this paper has been shown to be an attractive method for medical images, in particular for use in web applications. Due to its low complexity it is readily implementable, and its compression and decompression speeds outpace the standard compression methods available for the web. This low complexity does not come at the cost of utility and flexibility: advanced features necessary for medical imaging such as: high bit depths, multiple channels, and progressive decompression in both resolution and quality such as provided by the advance JPEG2000 compression method are available. A mild penalty in compression density is paid over JPEG2000 (3.4\% larger files with the medical images used for the benchmarks in this paper), yet compression ratio is still better than attained with the PNG compression format, which is the default choice of medical image data for web applications (or the uncompressed BMP) due to the poor support of JPEG2000 in web browsers today.

\section{Acknowledgments and Attributions}
Data used in this publication were generated by the National Cancer Institute Clinical Proteomic Tumor Analysis Consortium (CPTAC). The results published here are in whole or part based upon data generated by the TCGA Research Network: http://cancergenome.nih.gov/.



\end{document}